\documentclass[12pt,rmp]{revtex4}
\usepackage{graphicx}
\usepackage{amssymb}
\usepackage{epsfig}
\usepackage{psfrag}
\usepackage{epstopdf}
\def\intern#1{\relax}
\def\intern#1{\fbox{#1}}
\newcommand{\be}[1]{\begin{equation}\label{#1}}
\newcommand{\ee}{\end{equation}}

\newcommand{\fig}[1]{figure \ref{#1}}
\newcommand{\inten}[2]{#1$\times10^{#2}$W/cm$^2$}

\begin{document}
\title{Complex non-equilibrium dynamics in plasmas}
\author{Jan Michael Rost}
\affiliation{
Max Planck Institute for the Physics of Complex Systems, Dresden}

\maketitle

\section{Introduction}

In this contribution we will discuss two new forms of
strongly coupled plasmas which have become possible to create and
observe in the laboratory only  recently.  They exhibit a wealth of intriguing
complex behavior which can be studied, in many cases for the first
time, experimentally.  Plasmas, gases of charged particles,
are universal in the sense that certain
properties of complex behavior do only depend on ratios of
characteristic parameters of the plasma, not on the parameters themselves.
Therefore, it is of fundamental and far reaching consequence, to be
able to create and observe a strongly coupled plasma since its
behavior is paradigmatic for an entire class of plasmas.

We will illustrate this universality by discussing ultracold plasmas \cite{xxx07}
 and laser generated nanoplasmas \cite{sasi+06}.  They
exist at temperatures which are about 10  
 orders of magnitudes apart
and at densities which are more than 12 orders of magnitude different,
yet they share essential features of complex non-equilibrium dynamics \cite{popa+05,gesa+07}.

Before we describe these plasmas in detail we briefly recall the
prerequisites of complex behavior in the context of charged particle
dynamics, i.e., a special form of non-linear systems.

\section{Complex behavior in dynamical systems based on interacting 
particles}

Massive charged particles, such as electrons or ions, interact through
the Coulomb force, which is a two-body force, attractive for charges
of opposite sign and repulsive for charges of the same sign.  The
result is a seemingly simple dynamical system, governed by Newtonian
dynamics and classical equations of motion.  These equations of motion
are in their form identical to those which describe the dynamics of
stars and planets due to gravitation.  And in this context it has
been known for very long time that already the three-body problem
(sun, moon and earth, e.g.) is not analytically solvable and can even
give rise to chaotic behavior \cite{gu+92}.  However, regular
motion and purely chaotic behavior are both quite simple.  Complex
behavior emerges through an interplay of regular and chaotic dynamics,
one can say complex systems live at the ``edge of chaos''  
 \cite{riro+02}.  Complex
behavior for three particles does, loosely speaking, not occur in the
following situations:
\begin{itemize}
\item[(a)] If the motion of all three particles is mutually unbound,
they will, if at all, approach each other and escape again to infinity
after a single collision: this will not generate complex dynamics.
\item[(b)] If two particles are tightly bound to each other, relative
to the third one, the three-body system effectively decouples into a
two + one body system which also shows simple behavior.  

\item[(c)] As
aforementioned, under certain circumstances three (or more) particles
show completely chaotic behavior in the sense that the dependence on
initial conditions does not play any role after a very short time of
motion.  This, so called, ergodic behavior is also simple: It is in
its unpredictability reliably predictable.  A good analogy is the most
irrational of all irrational numbers, the golden mean.  It is most
irrational since its decimal representation has period one, i.e., one
cannot say which digit follows if the previous one is known.  On the
other hand, in the representation of a chain fraction, the golden mean
is the simplest number, given by $1/(1+1/(1+1/(1+\ldots$.
\end{itemize}

Hence, the parameters of the system must be such that (a)--(c) are
avoided, i.e., that the three (or more) particles are confined to a
certain volume and are forced to meet each other again and again
creating a history of collisions and eventually very complex behavior,
without loosing memory of the past which happens in a billiard game.
Here, a human analogy comes immediately to mind: Complex relationships
only develop through repeated encounters between persons entailing
memories of previous encounters. 
\section{Plasmas}
\label{sec:intro}
 Plasmas are a prototype of complex
systems in physics regarding property (b) as described in the last 
section. Innocently defined as a neutral gas of charged particles, they exhibit a wealth of phenomena which require the cooperate
if not collective motion of individual particles.  The best known
collective behavior is collective motion at specific frequencies,
where the fundamental one, the plasma frequency $\Omega$, depends only on the
density $\rho$, mass $m$ and charge $eZ$ of the particles, 
\begin{equation}
\Omega=\sqrt{4\pi e^{2}Z^{2}\rho/m}\,.
\end{equation}  Other, so called plasmon
modes, refer, e.g., to surface excitations and depend critically on
the geometrical shape of the plasma.

\begin{figure}[!htb]
\begin{center}
\includegraphics[width=0.8\textwidth]{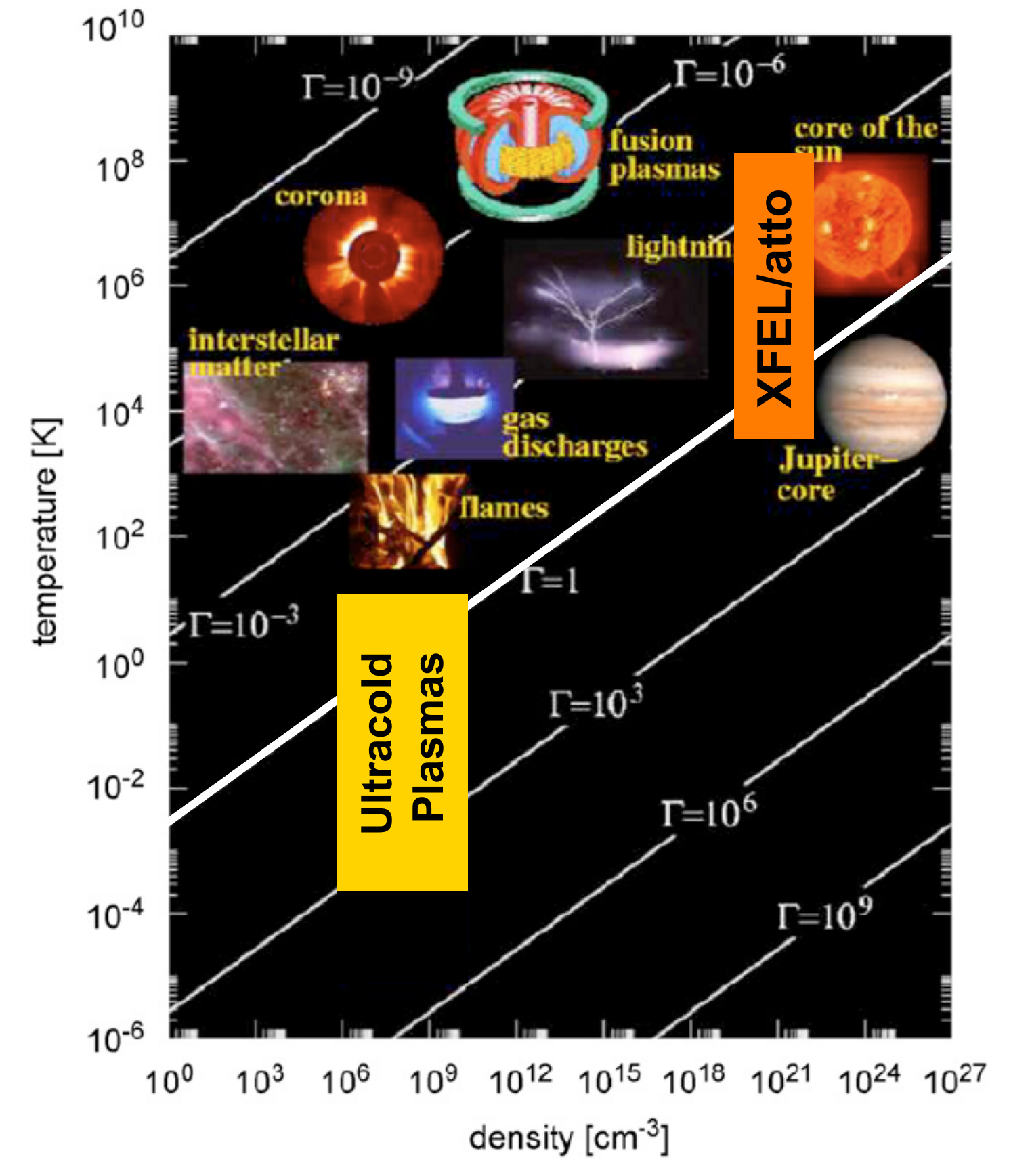}
\end{center}
\caption{Natural and man-made plasmas in the density- temperature
plane, after \cite{xxx07}. 
Lines of equal Coulomb coupling parameter $\Gamma$
(see text) are given in white, the thick white line corresponds to
$\Gamma=1$, the dividing line between ideal plasmas ($\Gamma\ll 1$)
and strongly coupled plasmas ($\Gamma\gg 1$).  Parameter ranges of the
plasmas discussed here are indicated by the yellow and orange bars for
ultracold and laser pulse generated plasmas, respectively.}
\label{fig:plasmas}
\end{figure}

Plasmas are abundant in the universe and therefore also called the
fourth state of matter.  The interior of Jupiter and
other giant stars consists of a plasma, so dense, that quantum
mechanical effects play a role.  More familiar plasmas are the corona
of the sun, and man made, magnetically confined plasmas for fusion
reactions.  An overview in term of temperature (kinetic energy of the
plasma) and density (potential energy) is given in \fig{fig:plasmas}.

From the view point of complexity, most of the plasmas in 
\fig{fig:plasmas},
however, are not so interesting since they fulfill property (c).
Being similar to ideal gases, they lack as an ensemble of particles
the relevance of memory regarding their collision history.  In this
limit, their Coulomb coupling parameter $\Gamma$ is very small.  It gives
the ratio between potential and kinetic energy in the plasma,
expressed through density (or inverse mean distance $a$ between
particles) and temperature $T$, 
\begin{equation}
\Gamma=e^{2}/(a\, kT)\,,
\end{equation}
while $k$ is the Boltzmann constant.
  On the other hand,
strongly coupled plasmas are characterized by dominant potential
energy, $\Gamma> 1$ and exhibit self organization such as
crystallization of the gas for $\Gamma>174$ \cite{duoc+99}, since the charged
particles tend to minimize their energy which requires to maximize
their mutual distances like in a crystal, if potential energy dominates.
One can see in \fig{fig:plasmas} that the few plasmas which are strongly coupled,
such as the core of Jupiter, are experimentally difficult to
access.

Most attempts to realize strongly coupled plasmas in the laboratory
have focused on dense matter.  With the advent of ultracold atomic
physics, the possibility to cool atoms in a trap to temperatures where
the gas even condenses into a Bose Einstein condensate, a new path to
strongly coupled plasmas has opened up: Ultracold plasmas \cite{xxx07}.
They are
extremely dilute with densities of less than $10^{10}$ particles per
cubic centimeter. Created from laser cooled atoms through photo 
ionization they are, however, so cold (in the micro-Kelvin regime for ions) that
despite their diluteness with large inter-particle separations,
potential energy wins over kinetic energy, allowing for $\Gamma>1$, 
see yellow bar in \fig{fig:plasmas}.  A
big advantage in terms of experimental controllability is the time
and spatial scale of these plasmas: The fundamental plasma frequency
(equation 1) reveals that they live on a
nanosecond/microsecond time scale for electrons/ions, respectively.
Spatially, the plasma clouds have a size of 1/10 mm, predefined by the
size of the traps.  These scales are convenient to experiment with.

Much more difficult to observe in the experiment due to the almost
solid state density is the second kind of plasma we are interested in: laser
generated nanoplasmas.  They live on a scale of 1 femtosecond ($10^{-15}$s) and have
extensions in the nanometer to sub-nanometer range, see orange bar in
\fig{fig:plasmas}.  Yet, recent advance in technology (attosecond ($10^{-18}$s)
laser pulses) will make it easier in the future also to observe these
strongly coupled plasmas.  While living in very different ranges
of space and time, laser generated plasmas share one property with
ultracold plasmas which is an important prerequisite for complex
behavior: specific geometric shapes due to confinement, either in a
trap or due to the form of the cluster, from which the nanoplasma forms.
This property of finite extension is important for complex behavior as
follows from conditions (a) and (c) above.

\section{Ultracold Plasmas}
\subsection{Plasma Creation}

\begin{figure}[!htb]
\begin{center}
\includegraphics[width=0.6\textwidth]{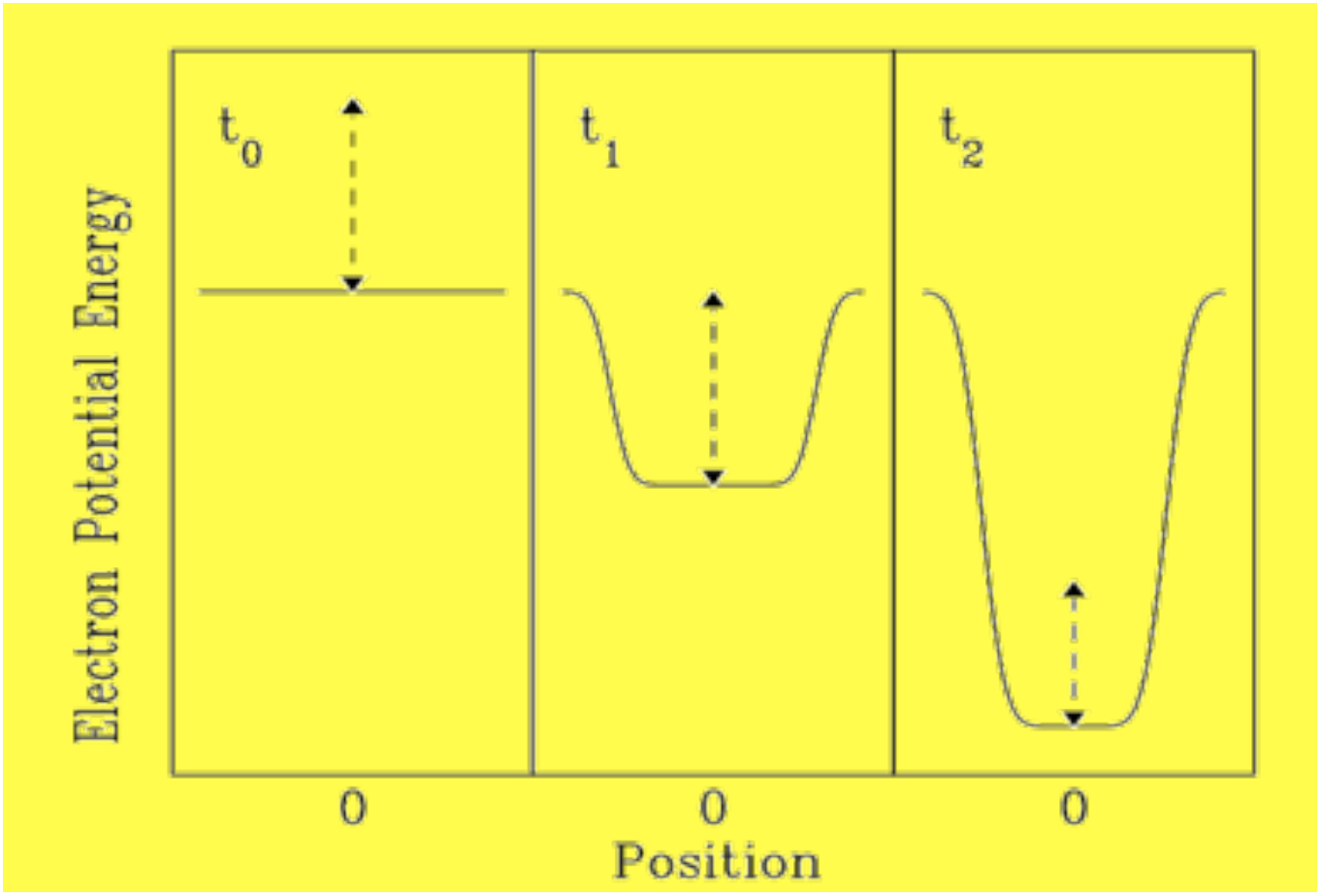}
\caption{Sketch of plasma formation by photo ionization of a geometrically confined gas, after  \cite{kiku+99}.
The left panels shows the neutral gas in the beginning at time $t_{0}$. Electrons ionized acquire a kinetic energy $E$ which is the difference between the photon energy $\hbar\omega$ and the atomic binding energy $I$ of the electron, $E=\hbar\omega-I$, indicated by the vertical dashed lines. The center panel captures the onset of plasma formation: due to multiple ionization a positive background potential of ions has formed which attracts the electrons. It is deep enough that the kinetic energy of a further electron ionized (indicated by an arrow) is not sufficient to escape from the cluster.
Right panel: photo ionization of electrons \emph{into} the plasma.}
\label{fig:plasmacreation}
\end{center}
\end{figure}

\begin{figure}[!htb]
\begin{center}
\includegraphics[width=0.7\textwidth]{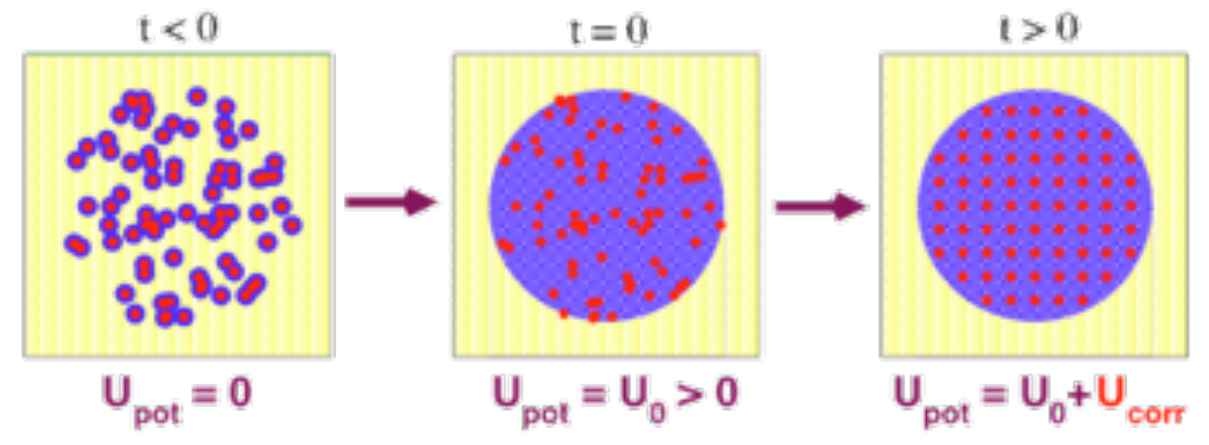}
\caption{A gas with randomly placed atoms (left panel, $t<0$) is ionized at time $t=0$.
As a consequence, there is potential energy $E_{\mathrm{pot}}=U_{0}$ between the ions. Subsequently, correlations develop by which the potential energy is lowered, the generated positive correlation energy $U_{\mathrm{corr}}$ appears as kinetic energy (higher temperature) of the ions ($t>0$).}
\label{fig:disheating}
\end{center}
\end{figure}

The creation of an ultracold plasma is sketched in
 \fig{fig:plasmacreation}.  First a gas of atoms is cooled in a magneto-optical trap.  Next,
the gas is excited to high Rydberg states or gently ionized just
above threshold.  As a consequence
electrons leave the gas and quickly a positive net background
charge of ions forms (the ions are so cold that they stand still on
this time scale) and eventually, the ionized electrons cannot leave
the gas anymore but form a plasma with very low kinetic energy.  At
the same time the ions also form a plasma whose Coulomb coupling
parameter is typically much higher due to the lower temperature of the ion plasma.
Formally, ionic Coulomb
Coupling parameters up to 10000 are possible.  This is too good to be
true, and indeed, it does not happen.  In fact, the laws of
thermodynamics teach us to expect due to equi-partitioning of energy
in the various degrees of freedom $\langle E_{\mathrm{pot}}\rangle\approx\langle
E_{\mathrm{kin}}\rangle$ and therefore $\Gamma=\langle
E_{\mathrm{pot}}\rangle/\langle E_{\mathrm{kin}}\rangle\approx1$.
This is indeed the case but now the question arises, where do the 
initially very cold ions $\langle 
E_{\mathrm{kin}}\rangle\approx0$ acquire so much energy? The reason is disorder 
induced heating sketched in \fig{fig:disheating}. The positions of the neutral 
atoms in the gas are random with a mean distance $a$ given by the 
density, $a\propto \rho^{-1/3}$.  After ionization the potential energy of 
the - now ions - is much higher than in an ordered, crystal like, 
state. Hence, through collisions, potential energy is converted into 
kinetic energy leading to $\Gamma\approx1$.
The process is quite violent and violates all laws of equilibrium
thermodynamics rendering it very interesting but beyond the scope of
our present considerations. Due to its  large size ($\approx 
100 \mu m$) and  slow dynamics ($\approx 1 \mu s$) ionic ultracold plasmas can be 
relatively easily monitored as described in  \cite{xxx07}.

The pertinent question, however, is: Can one realize a strongly coupled
ultracold plasma despite the disappointing heating effect? In principle, one could try to drive both components,
the electronic and the ionic one, to the strong coupling regime. However, since they evolve on a slower and experimentally therefore easier accessible time scale, we concentrate on the ions.  Our
proposal which we have underlined by calculations, consists of
additional laser cooling for the ions via an optical transition.  

\subsection{Cooling to the strong coupling regime}
Additional cooling of the ions in the plasma can be achieved
for strontium where the ion still has one optically active electron
while the first one has been fed into the plasma.  One question
remains, only the calculation or the experiment can answer: Is
there enough time after the creation of the ions to cool them down?
The cooling competes with the slow, but steady expansion of the plasma
which is no longer held in the trap.  Cooling can be modeled quite
reliably with a stochastic process \cite{popa+04b}, and the first
promising observation is that the expansion of the ions under cooling
slows down, not only quantitatively, but also qualitatively.
While the mean width $\sigma$ of the initially
Gaussian shaped cloud of the plasma in the trap (which remains
Gaussian through a selfsimilar expansion process) grows quadratically
without couling $\sigma \propto t^{2}$, this changes to a
long-time square-root dependence under cooling
$\sigma_{\mathrm{cool}}\propto \sqrt{t/\tau}$. Very important for the 
experimental feasability is the time scale $\tau = \beta m \sigma^{2}/(k T_{e})$. It can be 
controlled by the cooling rate $\beta$, the mass of the ion species 
$m$ and the initial width $\sigma$ and electronic temperature $T_{e}$ of the plasma.
Increasing any but the last parameter increases $\tau$ and therefore favors crystallization.
For better cooling with larger $\beta$ this is obvious, larger mass also makes the system to expand more slowly,
so does a larger initial size (at the same density) since it implies reduced repulsion of the ions which slows down the Coulomb explosion. Only increasing the initial electron temperature $T_{e}$  drives the ions more easily apart and should therefore be avoided.

\subsection{Self organized crystallization during expansion}
\begin{figure}[!htb]
\begin{center}
\includegraphics[width=0.45\textwidth]{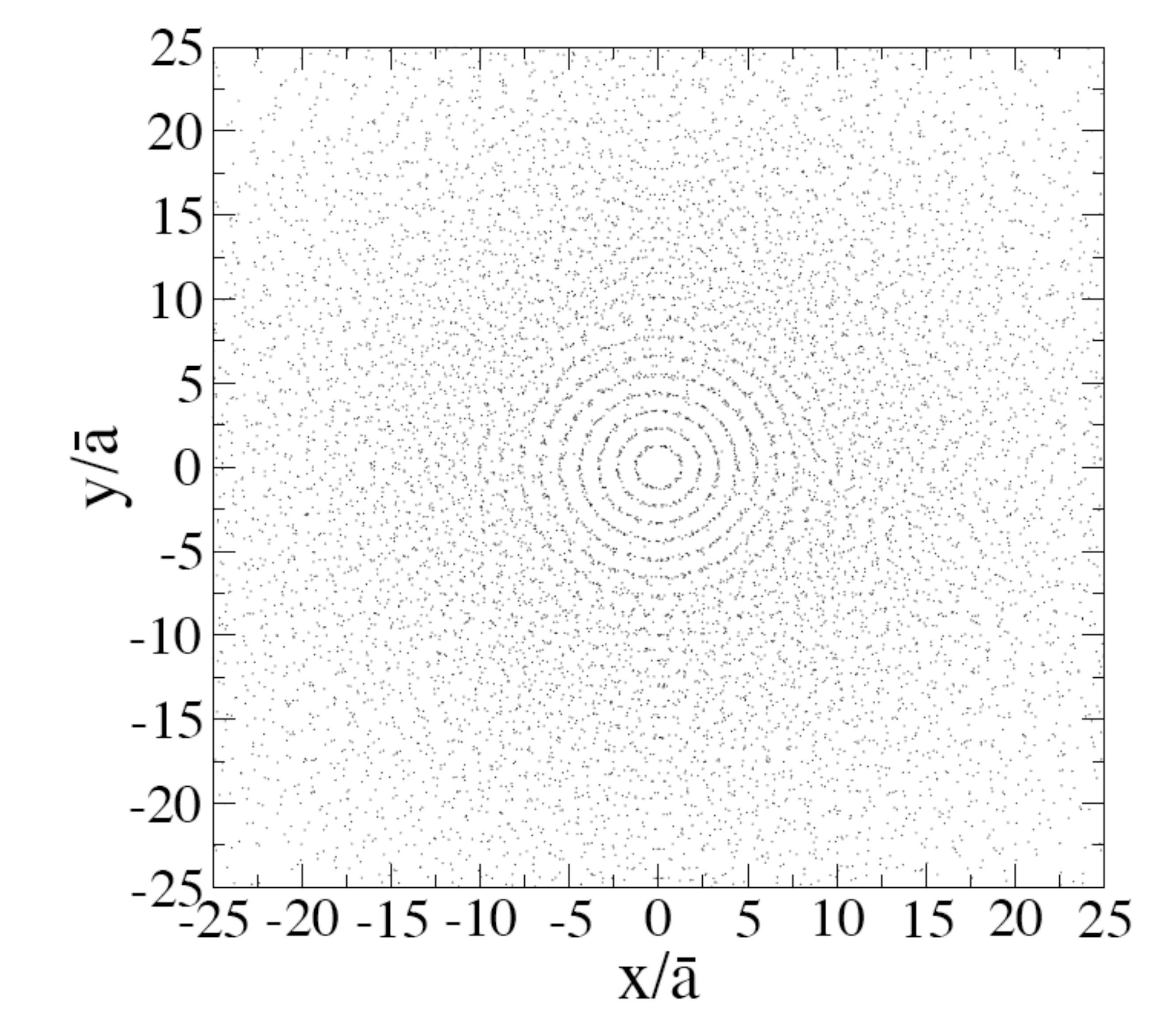}
\hfill\includegraphics[width=0.45\textwidth]{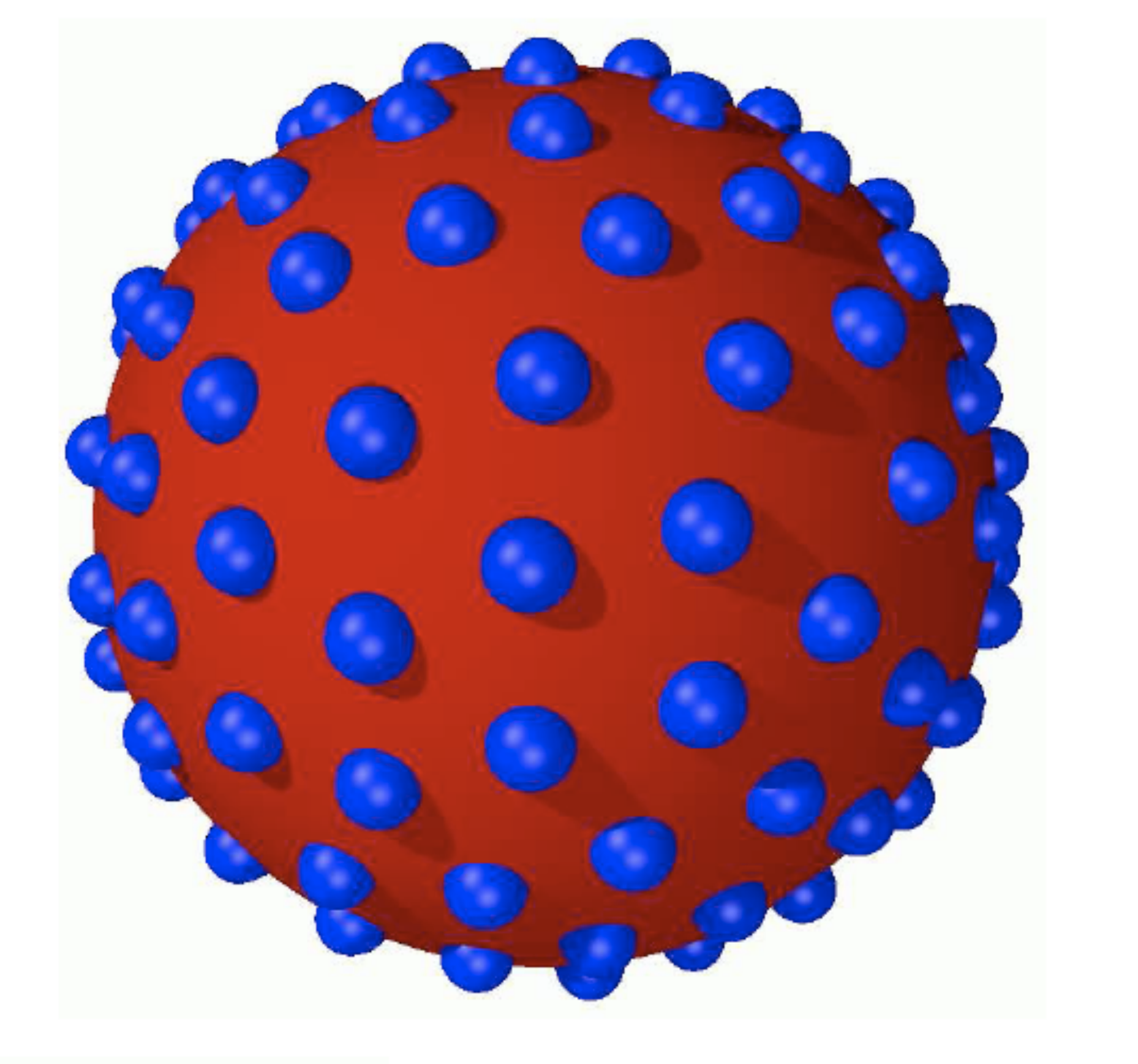}
\caption{Selforganization of a plasma of 80000 Beryllium ions  after  expansion time $t=216/\Omega$  into a crystal upon cooling with a rate of $\beta = 0.2\,\Omega$. The initial electronic Coulomb coupling parameter was $\Gamma_{e}=0.08$, after \cite{po+05}.
Shown is on the left a cut through the center of the plasma, with the length scaled by the local mean distance $a$ of ions. Distributions for several orientations have been superimposed for clarity of the structure.
 On the right the fourth shell of ions is shown (fourth circle from the center in the left figure).}
\label{fig:igel}
\end{center}
\end{figure}

What cooling achieves is to increase the Coulomb coupling parameter
$\Gamma$ which should lead to structure in the disordered gas of ions.
 Indeed, for sufficiently large $\tau$ the ions organize themselves on concentric
shells with no ions in between, i.e., they start to form an ion crystal 
during the expansion of the ion gas -- without any external force or
geometric boundary, see \fig{fig:igel}.

This crystallization is still a prediction which awaits experimental
verification facing two difficulties to overcome: first, sufficient
cooling, second, a detection scheme which can prove the crystallization
which is indeed somewhat more tricky.

However, another signature of complex behavior, the temporal 
evolution of the Coulomb coupling parameter, has already been measured and we 
will come back to it in section \ref{sec:NEplasmas}.

\section{Plasmas generated by short laser pulses}
\subsection{Plasma creation}

This type of plasma lives from the efficient deposition of energy to 
material in a very short time. Electrons absorb an enormous amount of energy from 
a very intense laser pulse lasting typically from a couple to 
some 100 fs and delivering $10^{14}-10^{18}$W/cm$^{2}$ power with a pulse 
profile in time close to a Gaussian. If the pulse is ramped up fast 
enough, electrons are ionized almost 
simultaneously, but certainly on the time scale of the respective 
plasma of a material of density $10^{22}\mathrm{cm^{-3}}$. This time 
scale is set by the electronic plasma frequency which is now of the order of a femtosecond.
The process of creating a plasma is very similar to the 
ultracold plasma (see \fig{fig:plasmacreation}), only much faster: While the first 
electrons ionized can leave, a positive background charge forms which 
further electrons cannot escape from since they do not have 
sufficient kinetic energy. When this happens, depends mostly on the 
laser frequency $\omega$ which determines through the energy of a 
single photon transferred  to an electron bound by energy $I$,  
how much kinetic energy $E$ is left to escape from the cluster with a 
(time-dependent)
background potential $V(t)$. 
 Plasma formation sets in when 
$E+V(t)=0$.

\begin{figure}[t-bp]
\begin{center}
\includegraphics[width=0.9\textwidth]{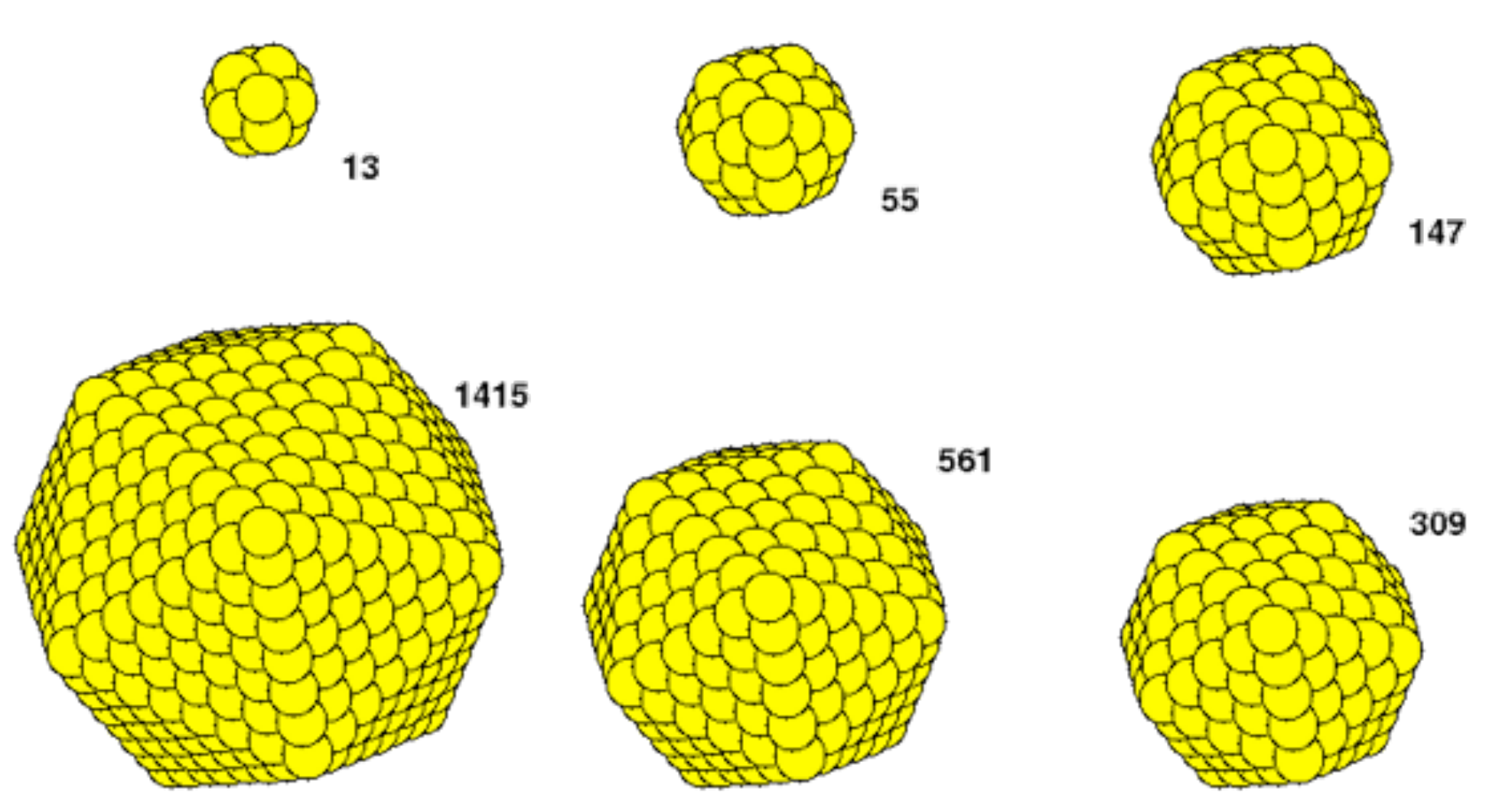}
\caption{Closed shell configurations of rare gas clusters form so 
called isocaeders and are particularly stable. They contain the 
number of atoms as indicated.}
\label{fig:geometries}
\end{center}
\end{figure}

We restrict ourselves here to clusters, i.e., droplets of atoms which
can exist from extremely small sizes (a few atoms) to almost
macroscopic sizes (micrometer diameter).  The relevant property in
terms of complex behavior is that they provide a geometric shape and
well defined confinement for the plasma (see \fig{fig:geometries}),
important for particles to undergo repeated encounters and thereby not
to loose memory of them (see conditions (b) and (c)).  Clearly, this
confinement is only of transient character since after multiple
ionization a cluster with repulsive ions will eventually undergo
Coulomb explosion.  Yet, on the plasma time scale, the cluster holds
together more than long enough for the complex plasma dynamics to
develop. Much more problematic is the detection of the plasma and of its 
evolution in time since this happens on a femtosecond scale.
\subsection{Detection of Ultrafast Plasma Dynamics}
We will discuss the scenario of a possible detection of the ultrafast
plasma dynamics in the context of laser pulses from unconventional
light sources.  One is a free electron laser  (FEL) at soft X-ray and soon
hard X-ray frequencies.  It is expected to deliver light pulses with
frequencies of $\omega = 12$eV$ - 12000$eV over less than 50fs and
with unprecedented intensities of more than $10^{20}$W/cm$^{2}$.  Light
pulses which such fantastic properties have their price: They are
generated by first accelerating electrons to almost the speed of light
over 2 km in a tunnel, then sending the electrons through a series of
magnets to force them on bent trajectories.  Thereby, they
radiate and organize themselves in bunches, leading overall to a
dramatic coherent amplification of the emitted light over a certain,
narrow frequency range.  This principle is called SASE (self amplified
stimulated emission) and machines working with it are built at DESY in
Hamburg and at SLAC in Stanford.

The second new development are attosecond pulses (1 as = $10^{-18}$s)
whose duration comes close to the period of about 100 as an electron
needs in the ground state of hydrogen to travel around the proton and
whose photon energy is between 20--150 eV. They are generated from
certain fractions of intense longer laser pulses at longer wavelengths,
and consequently, their intensity is much lower than the light from
FELs.  More details can be found in \cite{attoworld}.

Not only the detection itself is difficult, also the monitoring of the 
plasma must be synchronized with its creation, in other words one
needs to know when the clock started. These   
set-ups are called pump-probe experiments: The first ``pump'' pulse 
creates the dynamics one is interested in while the second ``probe'' 
pulse, probes the dynamics at a well defined time delay $\Delta t$.

\begin{figure}[!htb]
\begin{center} 
\includegraphics[width=0.9\textwidth]{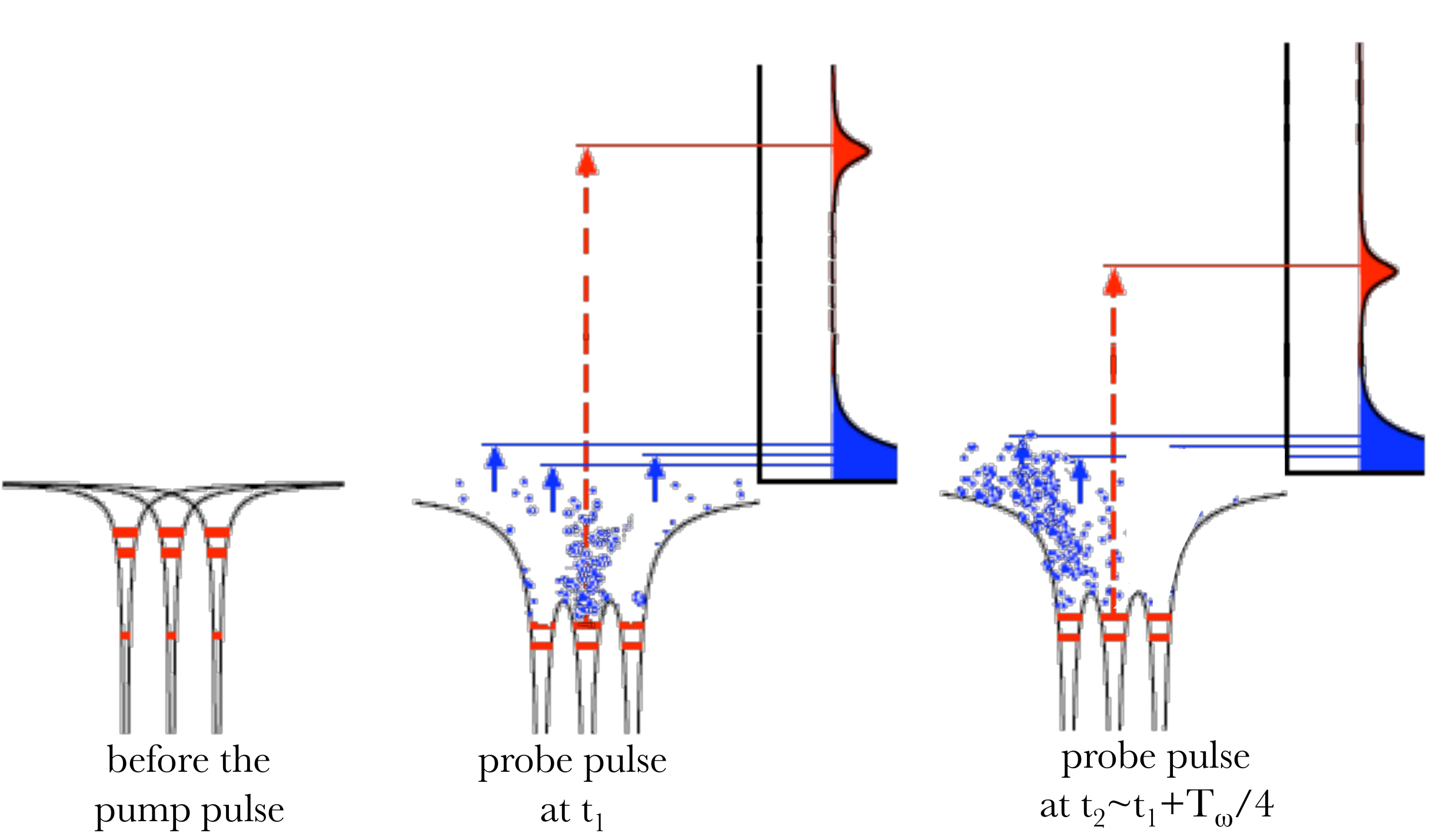}
\caption{Sketch of a double-pulse pump-probe measurement of electronic plasma dynamics, for details see text.}
\label{fig:pump-probe}
\end{center}
\end{figure}

\begin{figure}[!htb]
\begin{center}
\includegraphics[width=0.5 \textwidth]{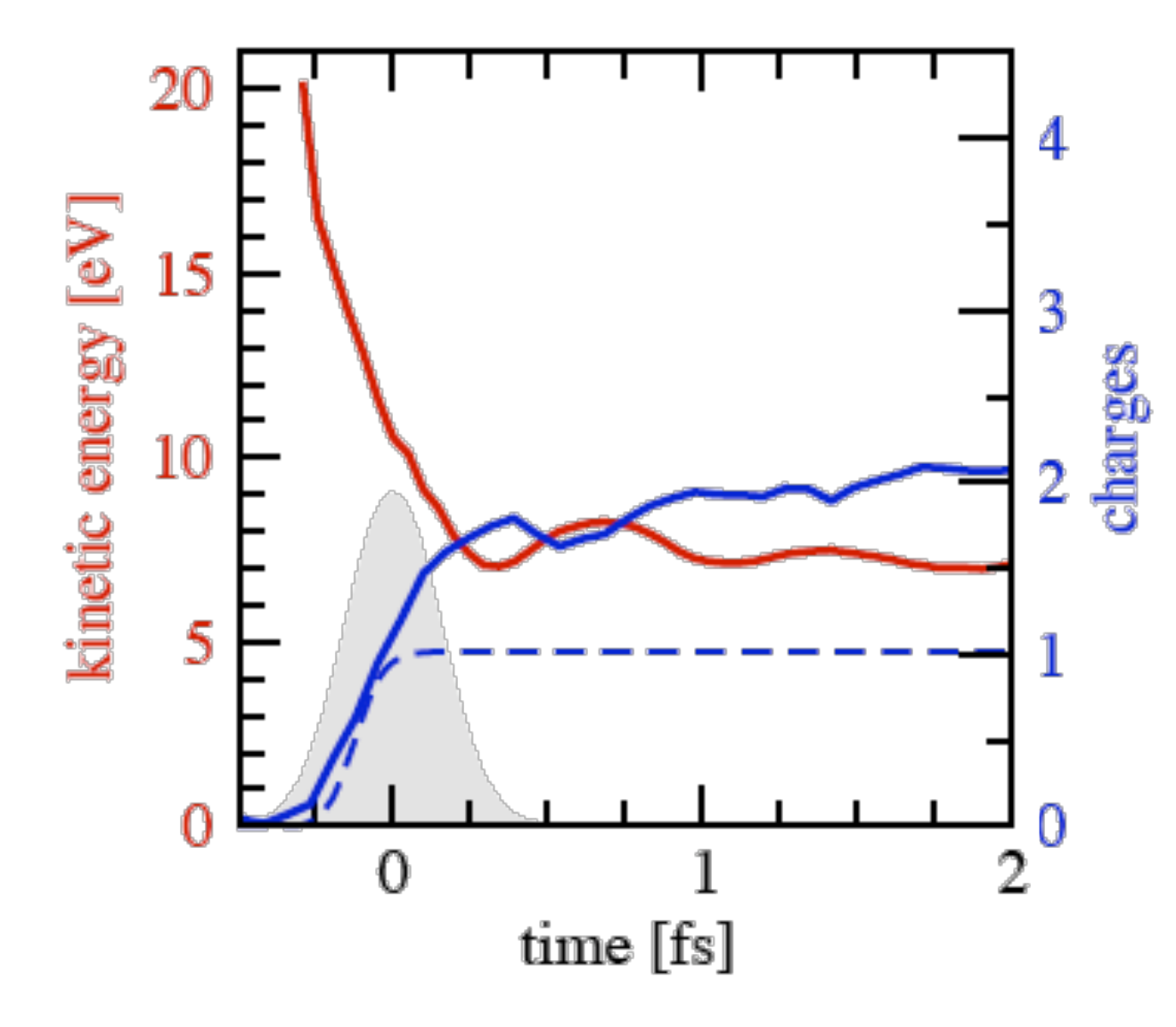}
\caption{Oscillation of the charges (black/blue, right axis) probed as in \fig{fig:pump-probe} with a pulse of variable time delay against the pump pulse (shown shaded, centered at time $t=0$) which triggers the plasma dynamics in
a cluster of 55 argon atoms, after \cite{sage+08}. The probed charges indirectly
map out the oscillation of the plasma electrons as is evident by the average kinetic energy of the plasma electrons (grey/red line, left axis), oscillating out of phase but with the same frequency.
 Also shown is the average charge state of the ions (dashed) which stays constant after the laser pulse is over. The two laser pulses have a photon energy of $\hbar\omega = 20$eV, peak intensity of \inten{5}{14} and a duration of 250 attoseconds.}
\label{fig:oscillation}
\end{center}
\end{figure}

A possible scenario is sketched in \fig{fig:pump-probe}.  Before the
first pulse hits (left panel of \fig{fig:pump-probe}), all electron
are bound to their mother atoms (the electronic levels are indicated
by dark vertical bars in the individual atomic potentials).  The pump
pulse removes the upper electrons in the atoms, and the potential
barriers between the ions are bent down.  Some of the electrons remain
in the cluster and form a plasma (small dots in the middle panel of
\fig{fig:pump-probe}).  A fraction of the distribution of plasma
electrons has kinetic energy high enough to be able to escape from the
cluster - they are measured as a continuous distribution (dark shaded)
in energy.  If the probe pulse is applied at a time $t_{1}$ and the
plasma electrons (which are assumed here to move collectively in a kind
of a wave) are in the center of the cluster, then electrons
ionized by the probe pulse from the dark (core) levels in the atoms are taken
to the continuum (dashed arrow) and they are measured at a certain energy
(light shaded peak).  If the probe pulse arrives when the plasma cloud
has passed the center of the cluster, then the electrons it ionizes
appear at lower energy in the continuum (light shaded peak in the
right panel of \fig{fig:pump-probe}).  The reason for the lower energy
 is that a time  $t_{1}$ the ionized electrons have to escape
from smaller positive background charge since the ions are screened
through the plasma electrons.  This is not the case at time $t_{2}$
and therefore, the ionized electrons loose more energy on the way to
the detector which becomes visible there. Hence, it is possible to
observe by ionization of core electrons at different probe times indirectly the oscillation of the plasma electrons as an oscillation of the kinetic energy of the ionized core electrons as shown in \fig{fig:oscillation}. 
The oscillation of the plasma electrons has a period of below 1 fs and can only be observed by laser pulses which are significantly shorter, in our simulation the pump and probe pulses last for 0.25 fs or 250 as.

\section{Non-equilibrium plasma dynamics}
\label{sec:NEplasmas}

The oscillation of the plasma electrons as described in the previous section is a quite complex collective response  of a strongly coupled plasma to a sudden perturbation: Driven out of equilibrium by the initial ionization and creation of the plasma within the 250 as of the pump pulse is too fast for  the system to adopt smoothly to this new situation. Rather, it
``overreacts'' which manifests itself in the oscillation of the plasma.  This oscillation damps out as the systems reaches equilibrium, but also due to loss processes.  

Why is this oscillation a signature of complex behavior?  It is organized, in the sense that the electrons move with a joint period, yet it is accidental since the electrons
are not required to have a specific initial condition for their motion, which is strongly influenced by collisions.

\begin{figure}[!htb]
\begin{center} 
\includegraphics[width=0.5\textwidth]{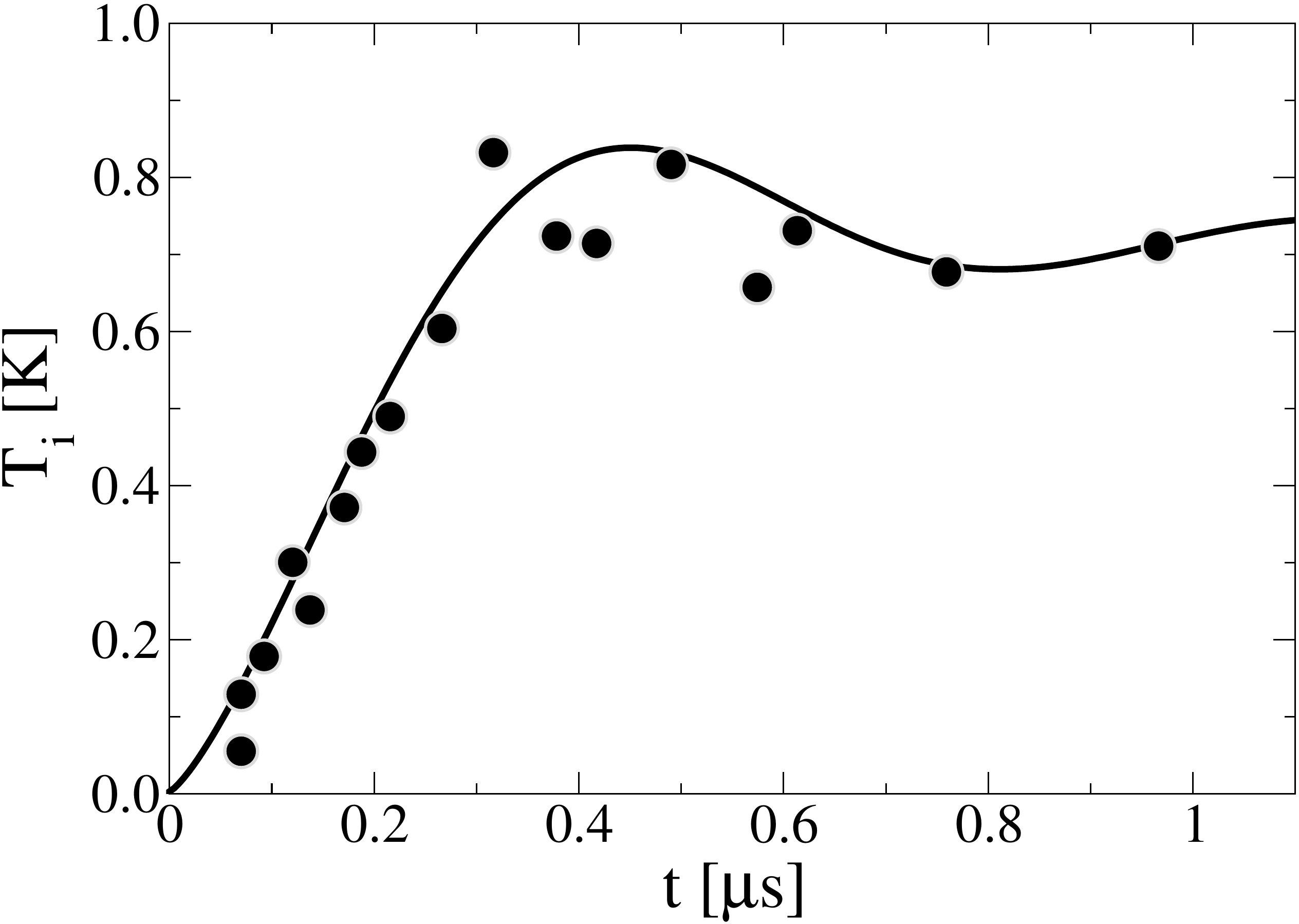}
\caption{Calculated  \cite{popa+05} 
(line) and measured \cite{chsi+04} 
(points) ion temperature of an expanding plasma of
1 million strontium ions with initial density $\rho = 2\times 10^{9}/$cm$^{3}$ and initial electron temperature of $T_{e}=38\,$K.}
\label{fig:gaoscillation}
\end{center}
\end{figure}

Even more fascinating, at least for a physicist, is the fact that this behavior is universal: It occurs in the same way for the ultracold plasma we have discussed previously (\fig{fig:oscillation}), but on a completely different, will say in comparison, extremely slow time scale of microseconds, see \fig{fig:gaoscillation}. Why is the ultracold plasma driven out of equilibrium by an excitation as slow as nanoseconds? Simply, because its response time scale due to the diluteness of the ions is microseconds, and on this time scale ionization and onset of plasma formation over a nanosecond is quite abrupt.  For the temperature, which is a measure for the kinetic energy of the (in this case ionic) plasma we observe exactly the same damped oscillatory behavior in time as in the case of the cluster. 

Hence, making use of the universality of the strong coupling phenomenon with complex collective behavior, depending only on the Coulomb coupling parameter being larger than unity, one can relatively easily see this behavior in the ultracold plasma with its
slow time scale compared to the laser generated plasma where the dynamics happens in complete analogy but extremely fast. However, even this fast motion will be accessible experimentally in the near future.

\section{Summary}

We have discussed universal complex behavior in two forms of microscopic plasmas, i.e., gases of charged particles. This complex behavior  manifests itself in an organized collective motion of the plasma particles.  The motion can happen on extremely different time scales, in our case the time scales are 10 orders of magnitude apart.
Complexity manifests itself since for this dynamics both, elements of order and chaos,
are neccessary in order to make it happen. This is a signature of complex systems which live on the boundary between regular and chaotic motion, colloquially speaking, near the ``edge'' of chaos.

\subsection*{Acknowledgment}
I am indebted to my collaborators, Thomas Pattard, Thomas Pohl, Ulf Saalmann and Ionut Georgescu for their valuable work and many discussions.

\end{document}